\def\etal{{\it et al.}}

\def\half{{\textstyle{1\over2}}}
\def\thalf{{\textstyle{3\over2}}}

\def\>{\rangle}
\def\<{\langle}

\def\rmb#1{{\bf #1}}

\def\beq{\begin{equation}}
\def\eeq{\end{equation}}

\def\beqy{\begin{eqnarray}}
\def\eeqy{\end{eqnarray}}

\def\l{\lambda} 
\documentclass[aps,showpacs,floatfix]{revtex4}
\newlength{\dinwidth}
\newlength{\dinmargin}
\begin{document}
\thispagestyle{empty}
\title{Strangeness -2 and -3 Baryons in a Constituent Quark Model\footnote{Notice: Authored by Jefferson Science Associates, LLC under U.S. DOE Contract No. DE-AC05-06OR23177. The U.S. Government retains a non-exclusive, paid-up, irrevocable, world-wide license to publish or reproduce this manuscript for U.S. Government purposes.}
}
\author{Muslema Pervin$^{1}$ and W. Roberts$^{2}$ }
\affiliation{$^1$ Physics Division, Argonne National Laboratory, Argonne, IL 60439\\
$^2$ Department of Physics, Florida State University, Tallahassee,
 FL 32306}

\begin{abstract} 
We apply a quark model developed in earlier work to the spectrum of
baryons with strangeness -2 and -3. The model describes a number of well-established
baryons successfully, and application to cascade baryons allows the quantum numbers of
some known states to be deduced.

\end{abstract}
\maketitle
\begin{flushright}JLAB-THY-07-728\end{flushright}

\setcounter{page}{1}

\section{Introduction and Motivation}

The study of hyperon properties can provide important insight to two questions
of crucial interest to hadronists. The first of these is `what are the relevant
degrees of freedom in a baryon', and is in some sense subsumed in the second,
`what is the mechanism of confinement?'. In order to understand the symmetries
and dynamics of the strong interaction, the expected multiplet structure of the
baryons must be established experimentally, and details of their excitation
spectrum are crucial. However, there has not been much information available on
hyperons, particularly those with $S<-1$, where $S$ is the strangeness of the
baryon. This means that neither the multiplets nor the excitation spectrum of
the light baryons are well established.

The most recent version of the Particle Data Group (PDG) listings \cite{Yao:2006px}
notes: {\it Not much is known about $\Xi$ resonances. This is because (1) they can only
be produced as a part of a final state, and so the analysis is more complicated than if
direct formation were possible, (2) the production cross sections are small (typically
a few $\mu$b), and (3) the final states are topologically complicated and difficult to
study with electronic techniques. Thus early information about $\Xi$ resonances came
entirely from bubble chamber experiments, where the numbers of events are small, and
only in the 1980âs did electronic experiments make any significant contributions.
However, nothing of significance on $\Xi$ resonances has been added since our 1988
edition.} Much of this comment is also valid for $\Omega$ baryons.

There are only three multistrange baryons with four star ratings whose spin and parity
are known, as reported in the most recent PDG listings~\cite{Yao:2006px}. These are the
ground states $\Xi(J^P=1/2^+)$ and $\Omega(3/2^+)$, and one excited cascade, 
$\Xi(3/2^+)$, with masses 1317 MeV, 1672 MeV and 1534 MeV, respectively. However, the
parity of the lowest lying $\Xi$ has not been determined experimentally, but positive
parity is expected. The spin and parity of the $\Omega^-$ have only been experimentally
determined  recently \cite{babar1}: the assignment of $J^P=3/2^+$ was based on
assignment of the state to the baryon decuplet. The PDG notes that for the $\Xi(3/2^+)$
at 1534 MeV, ``[s]pin-parity $3/2^+$ is favored by the data''. There are nine other
excited cascades and three other $\Omega$'s reported by PDG, but these all have
three-star or lower ratings, with few of their quantum numbers determined. Among the
three-star states, the $\Xi(1823)$ achieves a $J^P$ assignment of $3/2^-$ by virtue of
one experiment that favors $J=3/2$ but which cannot make a parity assignment
\cite{casc1}, and one experiment that determines that $J$ is consistent with 3/2 and
which favors negative parity \cite{casc2}. Among the other three-star states, it is
known that the state at 2030 MeV has $J\ge 5/2$ \cite{hemingway}, but the quantum
numbers of no other states have been ascertained. Among the $\Omega$ baryons, the only
state that has quantum number assignments is the ground state at 1672 MeV. The status
of the properties of these multistrange baryons is summarized in Table~\ref{hyperon}.

\begin{center}
\begin{table}[h]
\caption{The $\Xi$ and $\Omega$ baryons as listed in the most recent Particle Data Group Listings~\cite{Yao:2006px}
\label{hyperon}}
\vspace{5mm}
\begin{tabular}{|c|c|c|c|c|c|}
\hline
Experimental State& $J^P$ & PDG rating& Experimental State& $J^P$ & PDG rating\\ \hline
$\Xi(1317)$ & $1/2^+$ (expected) &****&$\Omega(1672)$ & $3/2^+$ (\cite{babar1}) &**** \\
$\Xi(1530)$ & $3/2^+$ (favored by data)&****&$\Omega(2250)$ & $?^?$ &*** \\
$\Xi(1823)$ & $3/2^-$ (\cite{casc1,casc2}) &*** &$\Omega(2380)$ & $?^?$  &** \\
$\Xi(1690)$ & $?^?$ &*** &$\Omega(2470)$ & $?^?$  &** \\
$\Xi(1950)$ & $?^?$ &*** &   &     &   \\
$\Xi(2030)$ & $\ge5/2^?$ (\cite{hemingway})&*** &   &     &   \\
$\Xi(2250)$ & $?^?$ &** &   &     &   \\
$\Xi(2370)$ & $?^?$ &** &   &     &   \\
$\Xi(1620)$ & $?^?$ &* &   &     &   \\
$\Xi(2120)$ & $?^?$ &* &   &     &   \\
$\Xi(2500)$ & $?^?$ &* &   &     &   \\\hline
\end{tabular}
\end{table}
\end{center}

Recently, after a hiatus of nearly twenty years, there has been increasing experimental
interest in hyperons with $S< -1$, due in part to the large samples of beauty and
charmed hadrons produced at Cleo and the $B$ factories, and the large number of
multi-particle decays made accessible by these samples. Results from BaBar
\cite{babar1,ziegler,Aubert:2006ux}, preliminary results from Jefferson Lab (JLab)
\cite{Price1,Price2}, and plans for experiments at JLab suggest that more
high-precision data on these states will be forthcoming in the not-too-distant future.
In addition, a number of activities related to these hyperons are being
carried out elsewhere around the world. For example, there have been measurements of
weak decays of cascades ($\Xi^0$) by the KTeV Collaboration~\cite{ktev}, as well as by
the NA48/I Collaboration~\cite{na48}, in addition to studies of cascade resonances in
relativistic heavy-ion collisions~\cite{Witt:2007xa}. 

The BaBar Collaboration at SLAC has been pursuing studies to measure the masses,
widths, spins and parities of a number of excited hyperons, including the $\Xi(1690)$.
The PDG gives this state a three-star rating, but its quantum numbers are undetermined.
In addition, the BaBar Collaboration has examined the $\Xi(1530)$ and $\Omega(1672)$ to
determine their quantum numbers. The result of these analyses is that a spin of 1/2 for
the $\Xi(1690)$ is better supported by experiment than spin 3/2 or 5/2, but the parity
remains undetermined. For the $\Xi(1530)$, the $J^P$ is determined as $3/2^+$, for the
first time. For the $\Omega(1672)$, the BaBar Collaboration concludes that the spin is
consistent with 3/2, if the decaying baryon has spin 1/2 (the processes studied are
$\Omega_c^0\to \Omega^- K^+$ and $\Xi_c^0\to\Omega^-\pi^+$) \cite{babar1,ziegler}.

On the theoretical side, there have been a few treatments of baryons with $S=-2$ and
$S=-3$. Within the framework of the constituent quark model, Chao, Isgur and
Karl~\cite{CIK} used a nonrelativistic quark model, while Capstick and Isgur~\cite{CI}
used a relativized version, both of which were based on one-gluon-exchange. Glozman and
Riska~\cite{Glozman,Glozman2} used a one-boson-exchange model to look at these states,
while Oh~\cite{Oh:2007cr} examined the hyperon spectrum in a Skyrme model. QCD sum
rules ~\cite{qcdsr} have also been used to examine these states, as has the collective
excitation model of Bijker, Iachello and Leviathan~\cite{BIL}. A number of authors have
also examined these states in the framework of the large $N_c$ expansion
\cite{largenc}.

All of these treatments describe the ground states of the $\Xi$ and the $\Omega$
spectrum successfully, but provide a range of predictions for the masses of the excited
states. For the lowest lying $\Xi(1/2^-)$, for instance, predicted masses range from
1550 MeV to 1869 MeV, with all but one of the approaches predicting masses larger than
1750 MeV. A spread of 100 to 200 MeV in the mass of any particular state is not
uncommon among the predictions of these various treatments, particularly for excited
states.

In the work we present in this manuscript, we use a nonrelativistic quark model to
obtain the excitation spectrum for multistrange baryons. In our calculation we fit a number
of the experimentally well-known baryons to fix the parameters of the model
Hamiltonian. We then use the same Hamiltonian to predict masses (and spin-parity
assignments) of a number of excited cascades and Omegas. As discussed above, an
outstanding question in the hyperon sector is that of which states belong to which
SU(6) multiplets, and this can only be  determined with any certitude if the spins and
parities of the states are known. The primary goal of this work is therefore to explore
the excitation spectrum of states with $S<-1$, and in doing so perhaps provide some
guidance to future experimental efforts that will examine these states. The model we
use is one that we have developed for examining semileptonic decays of
baryons~\cite{Pervin:2006ie, Pervin:2005ve}, and some details of the model are
presented in the next section. Section III presents our results, and section IV of this
manuscript provides our conclusions and describes possible future directions.

\section{The Model}

Our starting point is a nonrelativistic quark model Hamiltonian, similar to that used by Isgur and Karl~\cite{isgurkarl}, and described in our earlier work \cite{Pervin:2006ie, Pervin:2005ve}.

\subsection{Hamiltonian}

The phenomenological Hamiltonian we use takes the form
\beq
\label{hamil}
H=\sum_i K_i +\sum_{i<j}\left( V^{ij}_{\rm conf}+H^{ij}_{\rm hyp}\right) +V_{\rm SO}+ C_{qqq}.
\eeq
$K_i$ is the kinetic energy of the $i$th quark, and takes the form
\beqy
K_i= \left( m_i+\frac{p_i^2}{2m_i} \right).
\eeqy
The spin independent confining potential consists of linear and Coulomb
components, 
\beqy
V^{ij}_{\rm conf}= \sum_{i<j=1}^3\left({br_{ij}\over 2}-
{2\alpha_{\rm Coul}\over3r_{ij}}\right).
\eeqy
The spin-dependent part of the potential is written as 
\begin{eqnarray}
H^{ij}_{\rm hyp} =\sum_{i<j=1}^3\left[{2\alpha_{\rm con}\over 3 m_i 
m_j}{8\pi\over 3} \rmb{S}_i\cdot\rmb{S}_j\delta^3(\rmb{r}_{ij})
+ {2\alpha_{\rm ten}\over 3m_i m_j}{1\over {r}^3_{ij}}\left(
{3\rmb{S}_i\cdot\rmb {r}_{ij} \rmb{S}_j\cdot\rmb {r}_{ij}\over {r}^2_{ij}} 
-\rmb{S}_i\cdot\rmb{S}_j\right) \right],
\end{eqnarray}
which consists of the contact and the tensor terms, with $r_{ij}=\vert\rmb{r}_i-\rmb{r}_j \vert$. In this work we use a simplified spin-orbit potential that takes the form,
\begin{eqnarray}
V_{\rm SO} = \frac{\alpha_{\rm SO}}{\rho^2+\lambda^2} \frac{\rmb{L}\cdot\rmb{S}}{(m_1+m_2+m_3)^2}. 
\end{eqnarray}
In this expression, $L$ is the total orbital angular momentum and $S$ is the total spin of the baryon. We note that this form is not very sensitive to the internal structure of the baryon.

\subsection{Baryon Wave Function}

In our model, a baryon wave function is described in terms of a totally antisymmetric color
wave function, multiplying a symmetric combination of flavor, space and spin wave
functions. The symmetric flavor-spin-space part of the baryon wave function is written as $\Psi_{A}^S = \phi_{A} ({\rm flavor})\psi_{A}({\bf\rho}, {\bf\l})\chi_{A}({\rm spin}),$ where ${\bf \rho}= \frac{1}{\sqrt2}({\bf r}_1 - {\bf
r}_2)$, ${\bf\lambda} =\frac{1}{\sqrt6}({\bf r}_1 + {\bf r}_2 - 2{\bf r}_3)$
are the Jacobi coordinates.

The total spin of the three spin-$1/2$ quarks can be either $3/2$ or $1/2$.  The
spin wave functions for the maximally stretched state in each case are 
\begin{eqnarray}
\chi_{3/2}^S(+3/2)  &=& |\uparrow\uparrow\uparrow\rangle, \nonumber\\
\chi_{1/2}^\rho(+1/2)  &=& 
\frac{1}{\sqrt{2}}(|\uparrow\downarrow\uparrow\rangle - 
|\downarrow\uparrow\uparrow\rangle), \nonumber\\
\chi_{1/2}^\lambda(+1/2)  &=&- 
\frac{1}{\sqrt{6}}(|\uparrow\downarrow\uparrow
\rangle  + |\downarrow\uparrow\uparrow\rangle -2|
\uparrow\uparrow\downarrow\rangle),
\end{eqnarray}
where $S$ labels the state as totally symmetric, while $\lambda/\rho$ denotes
the mixed symmetric states that are symmetric/antisymmetric under the exchange
of quarks $1$ and $2$. In our model, we treat states with three identical quarks, such as the $N$, $\Delta$ and $\Omega$, differently from states with one distinguishable quark, such as the $\Lambda_Q$, $\Sigma_Q$ and $\Xi$.

\subsubsection{States with $m_{q_1} = m_{q_2} \ne m_{q_3}$}

For baryons containing one constituent quark having a mass different from that of the
other two quarks, we symmetrize or antisymmetrize the flavor-spin-space part of the wave function only with respect to interchange of the two identical quarks, denoted 1 and 2 in our model. The flavor part of the wave function for such states can be either symmetric or antisymmetric in quarks 1 and 2. For $\Lambda_Q$-type baryons the flavor wave
function is
\begin{eqnarray}
\phi_{\Lambda_Q} =  \frac{1}{\sqrt{2}}(ud - du)Q,
\end{eqnarray}
which is antisymmetric in quarks $1$ and $2$. The space-spin
portion of such wave functions must therefore be antisymmetric in quarks
$1$ and $2$, in order to yield a wave function that is symmetric in quarks 1 and 2.
The flavor wave function of a $\Sigma_Q$-type baryon is
\begin{eqnarray}
\phi_{\Sigma_Q} = \frac{1}{\sqrt{2}}(ud + du)Q,
\end{eqnarray}
which is symmetric in quarks $1$ and $2$. The space-spin
part of the wave function must therefore be symmetric in quarks
$1$ and $2$ to give the correct overall symmetry.

In either case, the spatial wave function for total
${\bf L}={\bf\ell}_\rho+{\bf\ell}_\lambda$ is constructed from a Clebsch-Gordan sum of the products of functions of the two Jacobi coordinates ${\bf \rho}$ and ${\bf \lambda}$, and takes the form 
\begin{eqnarray}
\psi_{LMn_{\rho}\ell_{\rho}n_{\lambda}\ell_\lambda}({\bf\rho}, {\bf\lambda}) = 
\sum_m\langle LM|\ell_{\rho}m,\ell_\lambda M-m\rangle\psi_{n_\rho \ell_\rho m}
({\bf \rho}) \psi_{n_\lambda \ell_\lambda M-m}({\bf \lambda}).
\end{eqnarray}
The spatial and spin wave functions are then coupled to
give wave functions corresponding to total spin $J$ and parity
$(-1)^{(l_\rho+l_\lambda)}$. Thus,
\begin{eqnarray}
\Psi_{JM}&=& \sum_{M_L}\< JM|LM_L, SM-M_L\>\psi_{LM_Ln_{\rho}
\ell_{\rho}n_{\lambda}\ell_\lambda}({\bf \rho}, {\bf 
\lambda})\chi_{S}(M-M_L)\nonumber\\
&\equiv&\left[\psi_{LM_Ln_{\rho}\ell_{\rho}n_{\lambda}\ell_\lambda}({\bf \rho}, {\bf \lambda})\chi_{S}(M-M_L)\right]_{J,M}.
\end{eqnarray}
The full wave function for a state $A$ is then built from a linear superposition of such components as 
\begin{equation}
\Psi_{A,J^PM}=\phi_A\sum_i \eta_i^A \Psi_{JM}^i.
\end{equation}
In the above, $\phi_A$ is the flavor wave function of the state $A$, and the expansion coefficients $\eta_i^A$ are determined by diagonalizing the Hamiltonian shown previously in the basis of the $\Psi_{JM}$. For this calculation, we limit the expansion in the
last equation to components that satisfy $N\le 2$, where
$N=2(n_\rho+n_\lambda)+\ell_\rho+\ell_\lambda$.
For example, wave functions for a $\Xi$ with $J^P=1/2^+$ have the form
\begin{eqnarray}\label{udsbasis}
\Psi^{\Xi}_{1/2^+M}&=&\phi_{\Xi}\left(\left[\vphantom{\sum_i}
\eta_1^{\Xi}\psi_{000000}({\bf \rho}, {\bf \lambda})
+\eta_2^{\Xi}\psi_{001000}({\bf \rho}, {\bf \lambda})
+\eta_3^{\Xi}\psi_{000010}({\bf \rho}, 
{\bf \lambda})\right]\chi_{1/2}^\lambda(M)\right.\nonumber \\
&+&\eta_4^{\Xi}\psi_{000101}({\bf \rho}, 
{\bf \lambda})\chi_{1/2}^\rho(M)
+\eta_5^{\Xi}\left[\vphantom{\sum_i}\psi_{1M_L0101}({\bf \rho}, {\bf \lambda})
\chi_{1/2}^\rho(M-M_L)\right]_{1/2, M}\nonumber  \\
&+&\left.\eta_6^{\Xi}\left[\vphantom{\sum_i}\psi_{2M_L0200}({\bf \rho}, {\bf \lambda})\chi_{3/2}^S(M-M_L)\right]_{1/2, M}
+\eta_7^{\Xi}\left[\vphantom{\sum_i}\psi_{2M_L0002}({\bf \rho}, {\bf \lambda})
\chi_{3/2}^S(M-M_L)\right]_{1/2, M}\right).
\end{eqnarray}
Diagonalization of the Hamiltonian yields seven states having $J^P=1/2^+$ composed of the
components above, with the set of $\{\eta_i^{\Xi}\}$ being different for each state.

\subsubsection{States with $m_{q_1} = m_{q_2} = m_{q_3}$}

Baryons containing three identical quarks have the full $SU(6)\equiv [SU(3)_{{\rm flavor}}\times SU(2)_{{\rm spin}}]$ symmetry. In our spectrum calculation we use the SU(6) symmetric wave functions, see for example Ref.~\cite{isgurkarl}, for such states. For completeness we give a very brief description of these SU(6) wave functions. Up to the $N=2$ harmonic oscillator level, there are five $SU(6)$ multiplets for positive parity baryons. In terms of the orbital quantum number $L$ these multiplets are the $56$ with $L=0$ and $L=2$,  the $70$, also with $L=0$ and $L=2$ and a $20$ with $L=1$. As an example, the predominant component of the ground state nucleon wave function is expected to be
\begin{eqnarray}
\left|^28 (56, 0^+) \frac{1}{2}^+\right\rangle= \frac{1}{\sqrt{2}}\left(\vphantom{\frac{1}{2}^+}\chi^\rho\phi^\rho+\chi^\l\phi^\l\right) \psi_{000000}({\bf \rho}, {\bf \lambda}),
\end{eqnarray}
where $\psi_{000000}({\bf \rho},{\bf \l})$ is the ground state $(l_\rho=l_\l=0)$ spatial wave function. The notation above is $\left|^{(2S+1)}\nu(\mu, L^P) J^P\right>$, where $S$ is the total spin of the quarks (= 1/2 or 3/2), $\nu$ is the SU(3)$_{\rm flavor}$ multiplet to which the state belongs, $\mu$ is its $SU(6)\equiv [SU(3)_{{\rm flavor}}\times SU(2)_{{\rm spin}}]$ multiplet, $L$ is the total orbital angular momentum in the state, $J$ is the total angular momentum, and $P$ is the parity. For the proton, 
\beqy
\phi^\rho=\frac{1}{\sqrt{2}}\left(ud-du\right) u, \,\,\,\,\phi^\lambda=
-\frac{1}{\sqrt{6}}\left[\left(ud+du\right)u-2uud\right].
\eeqy

As with the states containing two different flavors of quarks, the wave functions of the states with three identical quarks are constructed from a linear superposition of all wave functions (up to $N=2$) having the appropriate quantum numbers. Thus, for example, the wave functions for nucleons with $J^P=1/2^+$ are written  
\begin{eqnarray}\label{su6nucleons}
\Psi^{N}_{1/2^+M}&=&\left(\vphantom{\sum_i}
\eta_1^{N}\left|^28 (56, 0^+) \frac{1}{2}^+\right\rangle
+\eta_2^{N}\left|^28 (56', 0^+) \frac{1}{2}^+\right\rangle
+\eta_3^{N}\left|^28 (70, 0^+) \frac{1}{2}^+\right\rangle\right.\nonumber \\
&+&\left.\eta_4^{N}\left|^28 (70, 2^+) \frac{1}{2}^+\right\rangle
+\eta_5^{N}\left|^28 (20, 1^+) \frac{1}{2}^+\right\rangle\right),
\end{eqnarray}
with the set of $\{\eta_i^{N}\}$ being different for each state. The explicit forms for these SU(6)$\times$O(3) wave functions in terms of the flavor, space and spin wave functions that we use are
\begin{eqnarray}
\left|^28 (56', 0^+) \frac{1}{2}^+\right\rangle&=& \frac{1}{2}\left(\vphantom{\frac{1}{2}^+}\chi^\rho\phi^\rho+\chi^\l\phi^\l\right) \left(\vphantom{\frac{1}{2}^+}\psi_{001000}({\bf \rho}, {\bf \lambda})+
\psi_{000010}({\bf \rho}, {\bf \lambda})\right),\nonumber\\
\left|^28 (70, 0^+) \frac{1}{2}^+\right\rangle&=& \frac{1}{2\sqrt{2}}\left(\vphantom{\frac{1}{2}^+}\chi^\rho\phi^\rho-\chi^\l\phi^\l\right) \left(\psi_{001000}({\bf \rho}, {\bf \lambda})-\vphantom{\frac{1}{2}^+}\psi_{000010}({\bf \rho}, {\bf \lambda})\right)+
\frac{1}{2}\left(\vphantom{\frac{1}{2}^+}\chi^\rho\phi^\l+\chi^\l\phi^\rho\right) \psi_{000101}({\bf \rho}, {\bf \lambda}),\nonumber\\
\left|^28 (70, 2^+) \frac{1}{2}^+\right\rangle&=& \frac{1}{\sqrt{2}}\left[\phi^\rho\psi_{2M_L0101}({\bf \rho}, {\bf \lambda})+\frac{1}{\sqrt{2}}\phi^\l\left(\vphantom{\frac{1}{2}^+}\psi_{2M_L0002}({\bf \rho}, {\bf \lambda})-\psi_{2M_L0200}({\bf \rho}, {\bf \lambda})\right)\right]\chi^S,\nonumber\\
\left|^28 (20, 1^+) \frac{1}{2}^+\right\rangle&=& \frac{1}{\sqrt{2}}\left(\vphantom{\frac{1}{2}^+}\chi^\rho\phi^\l-\chi^\l\phi^\rho\right) \psi_{1M_L0101}({\bf \rho}, {\bf \lambda}).
\end{eqnarray}

For states with three identical quarks, as well as for states with only two identical quarks, we construct our wave functions using the harmonic oscillator basis. Each basis wave function takes the well-known form
\beq\label{hoa}
\psi_{nLm} ({\bf r})= \left[\frac{2\,n!}{\left(n + L +\half\right)!}
\right]^{\half} \alpha^{L+\thalf} 
e^{-\frac{\alpha^2r^2}{2}}
L_n^{L+\half}(\alpha^2r^2){\cal Y}_{Lm}({\bf r}),
\eeq
where ${\cal Y}_{Lm}({\bf r})$ is a solid harmonic, and $L_n^{\beta}(x)$ is a generalized Laguerre polynomial. The size parameters $\alpha_\rho$ and $\alpha_\lambda$ appearing in the wave functions are treated as independent variational parameters. However, when the three quarks in the baryon are identical, the variational procedure automatically chooses $\alpha_\rho = \alpha_\lambda$.

\section{Numerical Results}

\subsection{Hamiltonian Parameters and Baryon Spectrum}

In the previous section, we introduced the Hamiltonian we use to obtain the baryon spectrum. There are ten free parameters to be determined for
the baryon spectrum: four quark masses ($m_u=m_d$, $m_s$, $m_c$ and $m_b$),
and six parameters of the potential ($\alpha_{\rm con}$, $\alpha_{\rm tens}$, $\alpha_{\rm Coul}$, $\alpha_{\rm SO}$, $b$ and $C_{qqq}$), and these are determined from
a `variational diagonalization' of the Hamiltonian. The variational
parameters are the wave function size parameters $\alpha_\rho$ and
$\alpha_\lambda$ of Eq.~(\ref{hoa}). This variational diagonalization
is accompanied by a fit to a number of states in the known spectrum, which yields the `best' values for the parameters.  The
values we obtain for the parameters of the Hamiltonian are shown in
Table~\ref{parameter1}.
\begin{center}
\begin{table}[h]
\caption{Hamiltonian parameters obtained from the fit to a selection of known baryons. \label{parameter1}}
\vspace{5mm}
\begin{tabular}{|cccccccccc|}
\hline
 $m_\sigma$ & $m_s$  & $m_c$  & $m_b$  & $b$& $\alpha_{\rm Coul}$ 
 &$\alpha_{\rm con}$&$\alpha_{\rm SO}$ &$\alpha_{\rm tens}$ & $C_{qqq}$  \\
 (GeV)&(GeV)&(GeV)&(GeV)&(GeV$^2$)& && (GeV) &&(GeV)\\ \hline
0.2848& 0.5553&  1.8182 & 5.2019 &  0.1540 &$\approx 0.0$&1.0844 &0.9321& -0.2230 & -1.4204\\ \hline
\end{tabular}
\end{table}
\end{center}

We use three independent parameters for the strengths of the Coulomb, contact and tensor pieces of the potential. We believe that this is justified as
the Hamiltonian we use is viewed as completely phenomenological. In many of the fits we have obtained, we find that the strength of the Coulomb interaction was
consistently small, suggesting that, within this model, that interaction does not play a crucial role. We have also fixed the value of this coupling at 0.1 and 0.2 to investigate its effect on the other parameters and on the spectrum. When this is done, correlations among the parameters mean that they all change but no single parameter changes by more than a few percent. The spectrum also changes, with the masses of states shifting by up to 20 MeV, but with no significant degradation in the quality of the fit we obtain. Wave function size parameters also change by a few percent.

In contrast with the coupling constant for the Coulomb interaction, the coupling constant that results for the contact interaction is quite large, emphasizing the key role that this interaction plays in hadron spectroscopy. The tensor coupling also turns out to
be relatively large, suggesting that this interaction, and the mixings that it induces between harmonic oscillator substates, are very important ingredients
in the spectroscopy of baryons.

The string tension $b$ is slightly smaller than the the nominal value of 0.18, but is
larger than the value of 0.1425 GeV$^2$ recently obtained by Swanson and collaborators~\cite{swanson}
 in their treatment of charmonia. The quark masses we obtain are
somewhat smaller than in our previous work, and this can be traced to two sources.
First, in our previous work, we neglected the tensor interaction, and including it will
necessarily result in modifications to all of the fit parameters. Second, we are not
including the semileptonic decay rate of the $\Lambda_c$ baryon in this fit. In
Ref.~\cite{Pervin:2006ie} when this rate is included, it tended to push the quark
masses, particularly those of the up, down and strange quarks, to higher values. 

Perhaps the biggest influence in changing the values of the parameters has been the use
of the SU(6) wave functions for the nucleons, Deltas and Omegas. In our previous work
on the baryon spectra~\cite{Pervin:2006ie,Pervin:2005ve}, we used the so-called $uds$
basis flavor wave functions. In this basis, the flavor wave functions of the proton and
the $\Delta^+$ are both $uud$, and the two states are constructed out of the same set
of spin-space wave function components. For instance, for $J^P=\frac{1}{2}^+$, the full
wave functions for both states would be written as $uud$ multiplying the spin-space
components of Eq.~(\ref{udsbasis}). Diagonalization would then yield the wave
functions for seven states, five of which would be identified as nucleons, based on
their symmetry under exchange of the first two quarks, while the remaining two would be
identified as Deltas, as their wave functions would be fully symmetric under exchange
of any pair of quarks. One important point to note is that the nucleons and Deltas were
diagonalized together, so that they had the same wave function size parameters
($\alpha_\rho,\,\,\alpha_\lambda$), and the orthogonality of their wave functions arose
from the coefficients of the spin-space components. One drawback of this basis was that it led to 'spurious' states in the case of $\Omega$ baryons, states that were the equivalent of nucleons, and these had to be identified and removed from the spectrum.

In the present work, we use the full SU(6) wave functions for nucleons and Deltas (as
well as Omegas). This means that orthogonality of the wave functions arises from the
flavor component. The wave functions for the five $J^P=\frac{1}{2}^+$ nucleons that
arise in our model are shown in Eq.~(\ref{su6nucleons}), while the two Deltas with the
same spin and parity have wave functions

\beq\label{su6deltas}
\Psi^{\Delta}_{1/2^+M}=\left(\vphantom{\sum_i}
\eta_1^{\Delta}\left|^210 (70, 0^+) \frac{1}{2}^+\right\rangle
+\eta_2^{\Delta}\left|^410 (56', 2^+) \frac{1}{2}^+\right\rangle\right),
\eeq
with
\beqy
\left|^210 (70, 0^+) \frac{1}{2}^+\right\rangle&=& \phi^S\frac{1}{\sqrt{2}}\left(\vphantom{\frac{1}{2}^+}\chi^\rho\psi_{001000}({\bf \rho}, 
{\bf \lambda})+\chi^\l\psi_{000010}({\bf \rho}, {\bf \lambda})\right),\nonumber\\
\left|^410 (56', 2^+) \frac{1}{2}^+\right\rangle&=&\chi^S\phi^S
\frac{1}{\sqrt{2}}\left(\vphantom{\frac{1}{2}^+}\psi_{2M_L0200}({\bf \rho}, {\bf \lambda})+\psi_{2M_L0002}({\bf \rho}, {\bf \lambda})\right).
\eeqy
The sets of nucleons and Deltas are treated separately in the present work, so that
nucleons with a particular spin and parity do not necessarily have the same wave
function size parameters as Deltas with the same spin and parity.

It is instructive to compare the matrix element of one of the spin-dependent operators
calculated in the $uds$ and SU(6) bases, respectively, to demonstrate how the choice of
basis plays a role in guiding the values of the parameters. Consider, for instance, the
matrix element of  ${\bf S}_1\cdot{\bf S}_2$, evaluated using only the first component
of Eq.~(\ref{udsbasis}), or the first component of Eq.~(\ref{su6nucleons}). In either
basis, this component is the predominant component of the wave function of the
ground-state nucleon. In the $uds$ basis, one finds

\beq
\left<{\bf S}_1\cdot{\bf S}_2\right>=\left<\chi^\l\phi^\l\left|{\bf S}_1\cdot{\bf S}_2
\right|\chi^\l\phi^\l\right>=\frac{1}{4},
\eeq
while in the SU(6) basis, the result is
\beq
\left<{\bf S}_1\cdot{\bf S}_2\right>=\frac{1}{2}\left<\left(\chi^\l\phi^\l+\chi^\rho\phi^\rho\right)\left|{\bf S}_1\cdot{\bf S}_2
\right|\left(\chi^\l\phi^\l+\chi^\rho\phi^\rho\right)\right>=\frac{1}{2}\left(\frac{1}{4}
\left<\phi^\l|\phi^\l\right>-\frac{3}{4}\left<\phi^\rho|\phi^\rho\right>
\right)=-\frac{1}{4}.
\eeq
Since this particular matrix element plays a crucial role in the splitting between the
ground state nucleon and Delta, for instance, it should not be a surprise that the fit
parameters we obtain in this work differ significantly from those reported earlier.

Before discussing the results, we make one more comment about the parameters shown in
Table~\ref{parameter1}. One may argue that, in a model such as this, the parameters of
the model can not be reported with an accuracy of more than two or three significant
digits. However, we are attempting to fit a spectrum in which the masses of many states
are known to better than one MeV, representing a precision of better than one part in
several thousand in many cases. It is that precision that drives the precision to which
our parameters need to be reported. As an example, the spectrum that we obtain with
$m_u=0.28$ GeV and $b=0.15$ GeV$^2$ is significantly different from the spectrum we
obtain using the values shown in the table.

A selection of states from the baryon spectrum we obtain is shown in
Table~\ref{bspectrum}. For most states reported, the model provides a satisfactory
description, with most masses being reproduced to within 20 MeV. The exceptions are the
$\Sigma_b$ and $\Sigma_b^*$, but the mass splitting between these two states is well
reproduced. In comparison with this, the mass splittings among the states consisting of
light quarks, such as the $\Delta-N$ or $\Sigma^*-\Sigma$ mass differences, are
somewhat smaller than the experimental values. Nevertheless, these results give us some
confidence that, when applied to hyperons with strangeness -1 and -2, the
predictions will be reliable.

\begin{center}
\begin{table}[h]
\caption{Baryon masses in GeV obtained in the quark model we use. The
first two columns identify the state and its experimental mass, while 
the last column shows the masses that result from the model. If the experimental uncertainty in the mass of a state is less than 1 MeV, the uncertainty is not reported. Where appropriate, the masses shown are the average of different charge states. All masses are in GeV.
\label{bspectrum}}
\vspace{5mm}
\begin{tabular}{|l|c|l|}
\hline
State& Experimental Mass& Model\\ \hline
$N(1/2^+)$&0.938&0.970\\
$\Delta(3/2^+)$&$1.232\pm 0.001$&1.232\\\hline 
$\Lambda(1/2^+)$&1.116&1.103\\\hline
$\Sigma(1/2^+)$&1.189&1.210\\
$\Sigma(3/2^+)$&1.385&1.379\\\hline
$\Lambda_c(1/2^+)$&2.285&2.268\\
$\Lambda_c(1/2^-)$&2.595&2.625\\
$\Lambda_c(3/2^-)$&2.628&2.636\\
$\Sigma_c(1/2^+)$&2.455&2.455\\
$\Sigma_c(3/2^+)$&2.518&2.519\\\hline
$\Omega_c(1/2^+)$&$2.698\pm 0.003$&2.718\\\hline
$\Lambda_b(1/2^+)$&$5.624\pm 0.002$&5.612\\ \hline
$\Sigma_b(1/2^+)$&$5.812\pm 0.002$&5.833\\
$\Sigma_b(3/2^+)$&$5.833\pm 0.002$&5.858\\\hline
\end{tabular}
\end{table}
\end{center}

\subsection{The $\Xi$ States}

Our model results for a portion of the $\Xi$ spectrum is shown in
Table~\ref{bspectrum1}. These results are discussed in some detail in the subsections
below.

\begin{center}
\begin{table}[h]
\caption{The $\Xi$ and $\Omega$ spectra obtained in this work. The
first two columns identify the state and its experimental mass and uncertainty in the mass, while 
the last shows the masses that result from the model. The spins and parities shown in the first column are our quark model assignments. Masses are in GeV.
\label{bspectrum1}}
\vspace{5mm}
\begin{tabular}{|l|c|c|c|c|}
\hline
$J^P$ & \multicolumn{2}{c|}{$\Xi$} & \multicolumn{2}{c|}{$\Omega$}\\\hline
&Experiment& Model&Experiment&Model\\ \hline
$1/2^+$&1.317$\pm$ 0.001&1.325 & - & 2.175\\
&-&1.891 & - &2.191\\
&-&2.014& - & - \\\hline
$3/2^+$&1.532$\pm$ 0.001&1.520& 1.672 & 1.656\\ 
&-&1.934& - &2.170\\ 
&-&2.020& - &2.182\\ \hline
$5/2^+$&1.950$\pm$ 0.015&1.936& - &2.178\\
&-&2.025& - &2.210\\ \hline
$7/2^+$&2.025$\pm$ 0.005&2.035& - &2.183\\ 
&&2.148& - &-\\ \hline
$1/2^-$&1.690$\pm$ 0.010&1.725& - &1.923\\
&-&1.811& - &-\\\hline
$3/2^-$&-&1.759& - &1.953\\
&1.823$\pm$ 0.005&1.826& - &-\\\hline
$5/2^-$&-&1.883& - &-\\ \hline
\end{tabular}
\end{table}
\end{center}
\subsubsection{The $\Xi$ States with Known $J^P$}

There are only three $\Xi$ states with `known' spin-parity: $\Xi(1317)$ with
$J^P=\frac{1}{2}^+$, $\Xi(1530)$ with $J^P=\frac{3}{2}^+$ and $\Xi(1823)$ with
$J^P=\frac{3}{2}^-$. Table~\ref{bspectrum1} shows that our model reproduces the masses
of the $\frac{1}{2}^+$and $\frac{3}{2}^+$ states quite well. However, as with other
states consisting solely of light quarks, the splitting between these states is smaller
than the experimental value. The model prediction for the mass of the lowest
$\frac{3}{2}^-$ state is 1759 MeV, significantly smaller than the experimental value of
1823 MeV. The second $\frac{3}{2}^-$ is predicted to have a mass of 1826 MeV, very
close to the mass of the experimental state. It would seem natural, therefore, to
assign the experimental state to be the second $\frac{3}{2}^-$, and to suggest that the
lowest $\frac{3}{2}^-$ is yet to be found. We make such an assignment with a cautionary
note that models such as this often under predict the masses of negative parity states.
Indeed, in this model, the $S_{11}(1535)$ is predicted about 90 MeV too light. On the
other hand, the $\Lambda(1520)$ is relatively well described, with the model mass being
14 MeV greater than the experimental mass. For these states, inclusion of spin-orbit
interaction that is more sensitive to the internal structure of the state is very important.

\subsubsection{$\Xi(1690)$}

In addition to the ground state and the state with $J^P=\frac{3}{2}^+$, the existence
of the  $\Xi(1690)$ is relatively certain, and its mass is fairly well known. Its spin
and parity, however, have not yet been established. A recent study by the BaBar
Collaboration~\cite{ziegler,babar1,Aubert:2006ux} concludes that the spin is
consistent with 1/2, while there have been suggestions that it is a negative parity
state. In our model, we treat this state as having $J^P=\frac{1}{2}^-$, and the mass
that results with this assumption is 1725 MeV, 35 MeV heavier than the nominal
mass of the state. A more microscopic treatment of spin-orbit interactions, can be expected to drive this state to slightly lower mass. A number of other authors have
predicted larger masses for this state ~\cite{CI,CIK,largenc,BIL}. The level of
accuracy we have obtained here is comparable to that obtained with some of the other
baryon states we fit, so we claim that the spin-parity for the $\Xi(1690)$ state
predicted by the model is $\frac{1}{2}^-$.

\subsubsection{The 3-star $\Xi$ states}

The two other 3-star states known experimentally are the $\Xi(1950)$ and
$\Xi(2030)$. The spin-parity is not known for either of these states, but the
$\Xi(2030)$ state is reported as having $J^P\ge \frac{5}{2}^+$ \cite{hemingway}. There are a few model states that could be assigned to these two experimental states. In
Table~\ref{bspectrum1}, we tentatively assign the $\Xi(1950)$ a spin-parity of $\frac{5}{2}^+$, in which case the model mass of 1936 MeV is in good agreement with the experimental mass. This state is also consistent with the model state with
$J^P=\frac{3}{2}^+$ with a mass of 1934 MeV. Note that the PDG comments that
there may be more than one baryon in this mass region, which would be completely
consistent with our model predictions. For the $\Xi(2030)$, our model suggests
that it has $J^P=\frac{7}{2}^+$, and the model mass of 2035 MeV is consistent
with the experimental mass of 2025 MeV. The state is also consistent with the second $5/2^+$ state with a mass of 2025 MeV.

\subsubsection{$\Xi(1620)$}

The $\Xi(1620)$ is the only other $\Xi$ state with a mass below 2 GeV.
Experimental evidence for this state is very weak, with a number of searches
yielding null results \cite{nulls}. If this state exists, it poses a problem for the present
model, as the lightest positive parity excited model state that lies near this mass would be a predominantly radial excitation with a mass more than 250 MeV larger than the
nominal mass of this state. However, this kind of problem is one that has occurred in other sectors in models like this one, the most famous example of which is the Roper resonance with a mass of 1440 MeV: most quark models cannot produce a radial excitation that is as light as the Roper. Alternatively,  this state could be assigned to be one of the predicted negative parity states, such as the lightest $3/2^-$ model state. In this case, the model state is about 130 MeV heavier than the experimental state.  Whatever its parity, if the $\Xi(1620)$ exists, the present model, along with many others, would predict a mass that is too large by more than 100 MeV.

\subsubsection{$\Omega$ Baryons}

There is a single $\Omega$ that is known with any certainty. The lightest excited state
known has a mass of 2252 MeV, but neither its spin nor its parity has been established.
In our model, there are a number of states with masses near 2200 MeV. These include two
states with $J^P=\frac{1}{2}^+$ (with predicted masses of 2175, and 2191 MeV), three
excitations with $J^P=\frac{3}{2}^+$ (with predicted masses of 2170, 2182 and 2194
MeV), two $\frac{5}{2}^+$ states at 2178 and 2210 MeV, and a $\frac{7}{2}^+$ state at 2183
MeV. None of the masses of the negative parity states predicted in the model are sufficiently large for them to be considered as candidates for this experimental state.  More experimental information is certainly needed in this sector of the baryon spectrum.

\section{Conclusion and Outlook}

A nonrelativistic quark model was used to examine the 4 and 3-star known $\Xi$ states
and their possible spin-parity assignments. Recent experimental developments and
measurements~\cite{ziegler, babar1, Price1, Price2, ktev} have revived interest in
these states, which can provide a window into the mechanism of confinement, and the
relevant degrees of freedom in a baryon. Because the mass of the strange quark differs
significantly from those of the up and down quarks, this window provides a somewhat
different view from that provided by nonstrange baryons. The spin and parity of only a
few of the well-known $\Xi$ states are known, and based on the analysis in our model,
tentative spin and parity assignments have been made for a number of excited cascades.
We have found that the well-known states, $\Xi(1317)$, $\Xi(1530)$ and $\Xi(1823)$ are
quite well reproduced by our model, while specific $J^P$ assignments are made for the
3-star states with unknown spin-parity. Our model suggests that the $\Xi(1690)$ state should be assigned a $J^P$ of $\frac{1}{2}^-$, while the $\Xi(1950)$ state is consistent with a model state of mass 1.934 GeV having $J^P=\frac{3}{2}^+$, as well as a model state of mass 1.936 GeV and $J^P=\frac{5}{2}^+$. The $\Xi(2030)$ is also consistent with two model states, one having $J^P=\frac{5}{2}^+$ (2.025 GeV), the other with  $J^P=\frac{7}{2}^+$ (2.035 GeV). These predicted masses are in good agreement with the experimental ones. In the case of the $\Omega$ baryons, the sparse experimental data do not allow a meaningful comparison with the model.

Of course, an analysis of the masses alone is often not enough to make a definitive
quark model assignment for the states like these. Analysis of the electromagnetic and
strong couplings of these states to other cascades, as well as to hyperons containing a
single strange quark, would go a long way in helping to pin down which model states
correspond to which experimental states. Unfortunately, the data currently available do
not allow any kind of meaningful quantitative analysis.

The work presented here can be extended in a number of directions. The heavy baryons,
particularly the heavy cascades, are of particular interest, as activity at the $B$
factories has been providing new information on these states. In addition, these states
provide ideal testing grounds for ideas regarding diquark clustering in baryons and
heavy quark symmetries. The spectrum of multiply-heavy baryons is also of interest,
especially as the states found by the SELEX Collaboration have not been confirmed by
other searches.

\acknowledgments

This work is supported by the Department of Energy, Office of Nuclear Physics,  under
contracts no. DE-AC02-06CH11357 (MP) and DE-AC05-06OR23177 (WR). WR is grateful to the Department of Physics, the College of Arts and Sciences and the Office of Research at Florida State University for partial support. The authors are
grateful to V. Ziegler and J. Goity for useful discussions.

\newif\ifmultiplepapers
\def\beginpapers{\multiplepaperstrue}
\def\endpapers{\multiplepapersfalse}  
\def\journal#1&#2(#3)#4{\rm #1~{\bf #2}\unskip, \rm  #4 (19#3)}
\def\trjrnl#1&#2(#3)#4{\rm #1~{\bf #2}\unskip, \rm #4 (19#3)}
\def\baps{\journal {Bull.} {Am.} {Phys.} {Soc.}&}
\def\jap{\journal J. {Appl.} {Phys.}&}
\def\prl{\journal {Phys.} {Rev.} {Lett.}&}
\def\pl{\journal {Phys.} {Lett.}&}
\def\pr{\journal {Phys.} {Rev.}&}
\def\np{\journal {Nucl.} {Phys.}&}
\def\rmp{\journal {Rev.} {Mod.} {Phys.}&}
\def\jmp{\journal J. {Math.} {Phys.}&}
\def\rmm{\journal {Revs.} {Mod.} {Math.}&}
\def\jetp{\journal {J.} {Exp.} {Theor.} {Phys.}&}
\def\sjetp{\trjrnl {Sov.} {Phys.} {JETP}&}
\def\dokl{\journal {Dokl.} {Akad.} Nauk USSR&}
\def\spd{\trjrnl {Sov.} {Phys.} {Dokl.}&}
\def\tmf{\journal {Theor.} {Mat.} {Fiz.}&}
\def\snp{\trjrnl {Sov.} J. {Nucl.} {Phys.}&}
\def\hpa{\journal {Helv.} {Phys.} Acta&}
\def\yf{\journal {Yad.} {Fiz.}&}
\def\zp{\journal Z. {Phys.}&}
\def\anp{\journal {Adv.} {Nucl.} {Phys.}&}
\def\ap{\journal {Ann.} {Phys.}&}
\def\am{\journal {Ann.} {Math.}&}
\def\nc{\journal {Nuo.} {Cim.}&}
\def\etal{{\sl et al.}}
\def\pre{\journal {Phys.} {Rep.}&}
\def\pca{\journal Physica (Utrecht)&}
\def\prs{\journal {Proc.} R. {Soc.} London &}
\def\jcp{\journal J. {Comp.} {Phys.}&}
\def\pna{\journal {Proc.} {Nat.} {Acad.}&}
\def\jpg{\journal J. {Phys.} G (Nuclear Physics)&}
\def\fort{\journal {Fortsch.} {Phys.}&}
\def\jfa{\journal {J.} {Func.} {Anal.}&}
\def\cmp{\journal {Comm.} {Math.} {Phys.}&}

\end{document}